\documentclass{aa}
\usepackage{graphicx,epsfig}
\begin{document}

\newcommand{\ea}{{et al.}}
\newcommand{\beq}{\begin{equation}}
\newcommand{\enq}{\end{equation}}
\newcommand{\bfg}{\begin{figure}}
\newcommand{\efg}{\end{figure}}
\newcommand{\bfa}{\begin{figure*}}
\newcommand{\efa}{\end{figure*}}
\newcommand{\bea}{\begin{eqnarray}}
\newcommand{\ena}{\end{eqnarray}}
\newcommand{\dd}{{\rm{d}}}
\newcommand{\dg}{^{\rm{o}}}
\newcommand{\thetao}{{\theta_{\rm{o}}}}
\newcommand{\nuo}{{\nu_{\rm{o}}}}
\newcommand{\thetae}{{\theta_{\rm{e}}}}
\newcommand{\nue}{{\nu_{\rm{e}}}}
\newcommand{\re}{{r_{\rm{e}}}}
\newcommand{\rin}{{r_{\rm{in}}}}
\newcommand{\rout}{{r_{\rm{out}}}}
\newcommand{\const}{{\mbox{const}}}
\newcommand{\bmath}[1]{\mbox{\boldmath{${#1}$}}}

\def\lb#1{{\protect\linebreak[#1]}}

\def\apj{ApJ}
\def\asa{A\&A}
\def\mnras{MNRAS}
\def\phr{Phys. Rev.}
\def\phrl{Phys. Rev. Lett.}

\def\ltsima{$\; \buildrel < \over \sim \;$}
\def\gtsima{$\; \buildrel > \over \sim \;$}
\def\simlt{\lower.5ex\hbox{\ltsima}}
\def\simgt{\lower.5ex\hbox{\gtsima}}
\def\ebf{{\bf e}}
\def\Ebf{{\bf E}}
\def\dpar{{\bf \partial}}
\def\ka{{\rm K}$\alpha$}
\def\kb{{\rm K}$\beta$}
\def\ee{\`{e}}

\title{The Low X-Ray State of LS 5039\,/\,RX\,J1826.2-1450}
\author{A.\ Martocchia\inst{1}, C.\ Motch\inst{1}, I.\ Negueruela\inst{2} } 
\offprints{A.\ Martocchia, {\sf martok@quasar.u-strasbg.fr}}
\institute{
CNRS / Observatoire Astronomique de Strasbourg, 11 rue de l'Universit\'e, 
F--67000 Strasbourg, France
\and
Departamento de F\'{\i}sica, Ingenier\'{\i}a de Sistemas y Teor\'{\i}a
de la Se\~nal, 
Escuela Polit\'ecnica Superior, University of Alicante, Ap. 99, E--03080
Alicante, Spain 
}   
\date{Received... Accepted...}

\abstract{Recent {\it XMM-Newton} and {\it Chandra} observations of the high mass
X-ray binary LS\,5039\,/\,RX\,J1826.2-1450 caught the source in a faint X-ray state. 
In contrast with previous
{\it RXTE} observations, we fail to detect any evidence of iron line emission. 
We also fail
to detect X-ray pulsations. The X-ray spectrum can be well fitted by a simple powerlaw,
slightly harder than in  previous observations, and does not require the presence of any
additional disk or blackbody component. {\it XMM-Newton} data imply an X-ray photoelectric
absorption ($N_{\rm H} \sim 7 \times 10^{21}$\,cm$^{-2}$) consistent with optical reddening,
indicating that no strong local absorption occurs at the time of these observations. 
We discuss possible source emission mechanisms and hypotheses on the nature of the 
compact object, giving particular emphasis to the young pulsar scenario.
 }

\authorrunning{A. Martocchia et al.}

\titlerunning{The Low X-Ray State of LS 5039\,/\,RX\,J1826.2-1450}

\maketitle

        \section{Introduction}
        \label{sec:intro}

LS\,5039/RX\,J1826.2-1450 is a massive X-ray binary (HMXB) identified
in the {\it ROSAT} all-sky survey by Motch \ea\ (1997). The dynamical
parameters of this source, as well as the hardness of its
X-ray emission, are consistent with the compact object 
being either a neutron star (NS) or a black hole (BH). It may be accreting
directly from the companion's wind, but the presence of an accretion disk 
has not been excluded yet.  

Optical observations (Motch \ea\ 1997) allowed the identification of the
optical counterpart, a  
bright V = 11.2, O6.5V((f)) star (Clark \ea\ 2001).  The optical
photometric variability is 
very small ($<0.01$ mag, Mart\'{\i} \ea\ 2004) and optical colours
yield $E(B-V) \sim 1.26$ (Motch \ea\ 1997), which implies (according
to the empirical formulae in Predehl \& Schmitt 1995)
$N_{\rm H} \sim 7.2 \times 10^{21}$\,cm$^{-2}$.  The source distance
has been estimated at  about 3.1 kpc (Motch \ea\ 1997), a value confirmed 
by Rib\'o \ea\ (2002) and adopted throughout this work. 

The binary parameters were inferred spectroscopically for the first
time  by McSwain  \ea\ (2001): they found a period $P_{\rm orb} =
4.117\pm0.011$ d (now revised: $4.4267\pm0.0005$ d) 
and a high eccentricity  ($e = 0.48\pm0.06$: McSwain \ea, 2004). 
The mass function is quite low ($f(m) = 0.0017\pm0.0005 M_\odot$),  yielding 
a lowest acceptable inclination $i > 9^{\circ}$ and a compact object mass 
$M_{co} < 8 M_\odot$. O6.5V((f)) stars typically have $M 
\sim 36 M_\odot$  and $R \sim 10 R_\odot$ (Howarth  \& Prinja 1989).  

Interestingly, LS\,5039 is a runaway system, escaping from
the Galactic plane  with a total systemic velocity of $\sim 150$
km\,s$^{-1}$ and a perpendicular  
component greater than 100\,km s$^{-1}$ with respect to the plane
itself  (Rib\'o \ea\ 2002; McSwain  \& Gies 2002).  


First hints of radio emission from LS\,5039 were given by the VLA
(Mart\'{\i} \ea\ 1998);  
afterwards, a GBI-NASA monitoring campaign (Rib\'o \ea\ 1999) showed a
moderate radio 
variability, with a clear non-thermal spectral index and no  bursting
activity. More recently, VLBA observations at milliarcsecond scales
revealed persistent emission from radio jets (Paredes \ea\ 2000, 2002): 
therefore,  
the source is usually referred to as one of the few known {\it microquasars}, 
and is one of the very few radio-emitting massive
X-ray binaries. The VLBA map shows bipolar jets, emerging for at least
6 milli-arcseconds from  a central core.

The source may have been detected in the $\gamma$-ray band ($E>100$ MeV):
an  association with 
the {\it EGRET} source 3EG J1824-1514 has been proposed  (Paredes \ea\ 2000,
2002) and discussed  
in the framework of a model in which photons are up-scattered by the
relativistic  electrons 
in a cylindrical inhomogeneous jet (Bosch-Ramon \& Paredes 2004)\\

A broad emission line was seen in {\it Rossi-XTE} PCA data at $E_0
\sim 6.6$  keV 
(Rib\'o \ea\ 1999), compatible with a slightly blue-shifted,
fluorescent  neutral 
iron feature, or with \ion{Fe}{xxv} emission at rest velocity. This
detection is of utmost 
interest, since fluorescent iron line emission can be used as a
powerful diagnostic
to test assumptions on the accretion flow and/or the jet;  however,
the PCA energy resolution is insufficient to distinguish between the
possible models for the line. 
Also for this reason, observations of LS\,5039\,/\,RX\,J1826.2-1450
with  the new 
generation X-ray satellites are of great importance. 

In the present work we therefore concentrate on recently obtained {\it
Chandra} and 
{\it XMM-Newton} data. After summarising the results obtained  with
previous X-ray 
missions (next Section), we describe the new data
(Section~\ref{sec:newdata}), and discuss the implications in the last Section. 

\begin{table}[htbp]
\centering
\begin{tabular}{|c|ccc|}
\hline
{\it Observation}       & $F_{0.3-10 keV}$ & Orbital          &   $\Gamma$     \cr
                        &                  & Phase            &                \cr
\hline
{\it RXTE} 08/02/98 (I) & $\sim 40$      & $\sim 0.91$ & $1.95\pm0.02$  \cr
{\it RXTE} 08/02/98 (II)& $\sim 40$      & $\sim 0.10$ & $1.95\pm0.02$  \cr
{\it RXTE} 16/02/98     & $\sim 40$      & $\sim 0.88$ & $1.95\pm0.02$  \cr
{\it ASCA} 04/10/99     & $\sim 13$      & 0.22-0.38   & $\sim 1.5$      \cr
{\it SAX} 08/10/00      & $\sim 4.9$     & 0.75-0.96   & $\simlt 1.8$   \cr
{\it Chandra} 10/09/02  & $\sim 8.1$     & $\sim 0.35$ & $1.15^{+0.23}_{-0.21}$  \cr
{\it XMM} 08/03/03      & $\sim 10.3$    & $\sim 0.79$ & $1.56^{+0.02}_{-0.05}$  \cr
{\it XMM} 27/03/03      & $\sim 9.7$     & $\sim 0.21$ & $1.49^{+0.05}_{-0.04}$   \cr
\hline
\end{tabular}
\caption{X-ray unabsorbed flux in the 0.3--10\,keV band (in units of $10^{-12}$\,erg 
cm$^{-2}$ s$^{-1}$), orbital phase and spectral powerlaw index of LS\,5059 in the 
observations performed with different satellites. 
The references for the {\it RXTE} and {\it BeppoSAX} observations are 
Rib\'o \ea\ (1999) and Reig \ea\ (2003), respectively; however, we checked PDS data 
ourselves to better constrain $\Gamma$ in the {\it BeppoSAX} observation (see text). 
We also analyzed {\it ASCA} SIS and GIS data,
obtaining the results which are reported here. Orbital phases are computed    
according to the revised ephemeris of McSwain et al. (2004); the associated errors range
from $2$ to $5 \%$. }
\label{gamma}
\end{table}
      
        \section{The X-Ray History of RX\,J1826.2-1450}
        \label{sec:xrays}

A first detection of LS\,5039's X-ray emission was given by {\it
ROSAT} (Motch \ea\ 1997),  which found the source at a luminosity
level $L_{\rm X} \sim 8.1 \times 10^{33}$ erg s$^{-1}$  in the 0.1--2.4 keV interval. 

{\it Rossi-XTE} ASM data, covering the period 1996 February -- 1998
November,  and PCA data 
obtained in 1998 were analysed by Rib\'o \ea\ (1999).  They found no
evidence of any X-ray 
periodicities (on timescales 2-200\,d and 0.02--2000\,s,
respectively), nor of any significant 
variation on the two-year timescale. PCA data show  the source at a
luminosity level of  
$\sim 6 \times 10^{34}$ erg s$^{-1}$ in the 3--30 keV interval
($F_{0.3-10 keV} \sim 40 \times 10^{-12}$ erg cm$^{-2}$ s$^{-1}$, unabsorbed). 
On this occasion, the spectrum was well described
by a bare powerlaw continuum  with $\Gamma \sim 1.95$, plus a strong (EW $\sim 0.75$\,keV)
and broad  ($\sigma \sim 0.39$ keV) Gaussian line feature at $6.62$\,keV, i.e. consistent
with a  multi-component or with a relativistically broadened iron line. Significant 
emission, with no exponential decay, was seen up to 30\,keV. 

More recently, in October 2000, LS\,5039 was observed for 80 ks with
{\it BeppoSAX} by Reig \ea\ (2003), who found a moderate absorption
($N_{\rm H} \sim 1 \times 
10^{22}$ cm$^{-2}$); however, no iron line was detected,  possibly due to the 
limited signal-to-noise. At the time the source was at a flux level  
$F_{0.3-10 keV} \sim 4.9 \times 10^{-12}$ erg cm$^{-2}$ s$^{-1}$ (unabsorbed,
corresponding to $L_{\rm X} \sim 5.6  \times 10^{33}$ erg s$^{-1}$
in the 1--10 keV band), almost one order of  magnitude lower than in the {\it
Rossi-XTE} observation performed two and a half years before. 
This negative trend is confirmed also looking at archival 1999 {\it ASCA} 
SIS and GIS data (also mentioned by Reig \ea, 2003), in which the source has  
$F_{0.3-10 keV} \sim 13 \times 10^{-12}$ erg cm$^{-2}$ s$^{-1}$. 
We re-analyzed these data and found they are best fitted by
a powerlaw with $\Gamma \sim 1.5$ (see Table~\ref{gamma}). 

The {\it BeppoSAX} LECS and MECS continuum was apparently best fitted by a
steeper powerlaw ($\Gamma \sim 1.8$). 
In general, much better constraints on the powerlaw index and, 
possibly,  cutoff energy, are given by the higher-energy {\it BeppoSAX} camera PDS,
which is very  well suited to that aim. Unfortunately, the PDS
collected too few counts during that observation to provide really
useful data. However, the general 
high-energy spectrum is  compatible with a harder powerlaw component
($\Gamma < 1.8$, our analysis). 
On the base of the orbital solution of McSwain \ea\ (2001), the covered
phase interval seemed favourable to observe an eclipse of the X-ray 
source by the primary star; the revised ephemeris (McSwain \ea\ 2004)
rather indicates that the {\it BeppoSAX} observations do not cover
periastron, but extend only till slightly before
phase 0 (cp. Table~\ref{gamma}). 

The overall flux history of LS\,5039\,/\,RX\,J1826.2-1450 is summarized 
in Table~\ref{gamma}.

        \section{Chandra and XMM-Newton observations}
        \label{sec:newdata}

In the following we report on recent {\it Chandra} and {\it XMM-Newton} observations of
LS\,5039. Data  reduction has been performed with the standard CIAO 3.0 and SAS 6.0 packages,
respectively, while for the fitting procedure we used XSPEC 11.2. We tried several spectral
models, listed below:

I. {\sc wabs * powerlaw } ;

IIa. {\sc wabs * ( powerlaw + gaussian )} , with $\sigma_{\rm line} \equiv 0.02$ keV;

IIb. {\sc wabs * ( powerlaw + gaussian )} , with $\sigma_{\rm line} \equiv 0.39$ keV
like in Rib\'o \ea\ (1999) ;

IIIa. {\sc wabs * ( diskbb + gaussian )} , with $\sigma_{\rm line} \equiv 0.02$ keV;

IIIb. {\sc wabs * ( diskbb + gaussian )} , with $\sigma_{\rm line} \equiv 0.39$ keV
like in Rib\'o \ea\ (1999) ;

IV. {\sc wabs * ( powerlaw + bbodyrad )} .

The results of the fits are shown in Table~\ref{tab:fits}. Unless stated otherwise, errors
are given at the 90 \% confidence level. We also tried a model similar to model IV replacing
the blackbody by a disk blackbody for the {\it XMM-Newton} data (see text below).

        \subsection{{\it Chandra}}
        \label{sec:chandra}

LS\,5039 was observed with {\it Chandra} ACIS on 2002, September 10, for about 10 ks 
(06:44:00--10:05:10 UT) in Faint Mode, through the High Energy Transmission Grating  (HETG).
We analysed the zeroth-order image, with slightly more
than 1000 source counts registered, corresponding -- when a simple
powerlaw continuum is assumed  (model I) -- to an unabsorbed 
flux level of $\sim 6.8 \times 10^{-12}$ erg cm$^{-2}$ s$^{-1}$  in the 2 to 10 keV band, 
equivalent to $\sim 8.1 \times 10^{-12}$ erg  cm$^{-2}$ s$^{-1}$ in the 0.3 -- 10
keV interval. This is slightly more than in the 2000 {\it BeppoSAX} pointing, which therefore
corresponded to the lowest observed state of this source. 
However, the spectrum is the hardest
in the {\it Chandra} observation ($\Gamma \sim 1.1$, see Table~\ref{gamma}).

We rebinned the ACIS spectrum to a minimum of 25 counts/bin and tried
to fit the data by adding 
an iron line and thermal disk emission (see Table~\ref{tab:fits}). Although the 
signal-to-noise ratio of the observation is rather poor, we can set some
constraints on individual 
spectral components: in particular, a narrow ($\sigma=20$ eV)  or
broad ($\sigma=390$ eV) Fe 
line is always compatible (within the 90~\% confidence level)
with having a null EW, 
and the F-test shows that its introduction never gives a significantly
better fit.  
 
Parameters cannot be really constrained in the powerlaw-plus-blackbody model. The
statistics are also too poor to extract any useful information on short-time variability
and/or X-ray pulsations.

        \subsection{{\it XMM-Newton}}
        \label{sec:newton}

\begin{figure*}[t]
\includegraphics[angle=-90,width=0.49\textwidth]{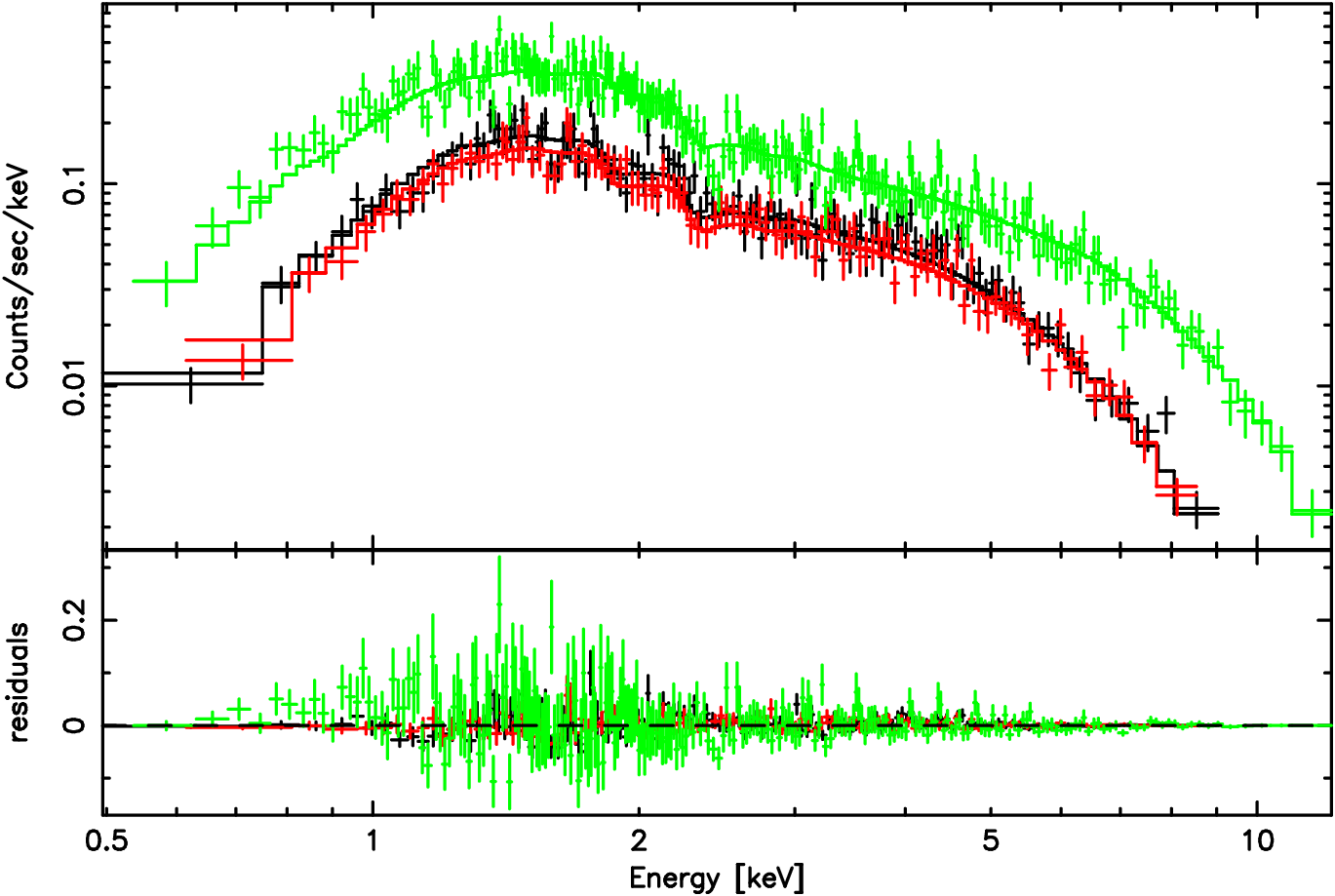}
\hfill
\includegraphics[angle=-90,width=0.49\textwidth]{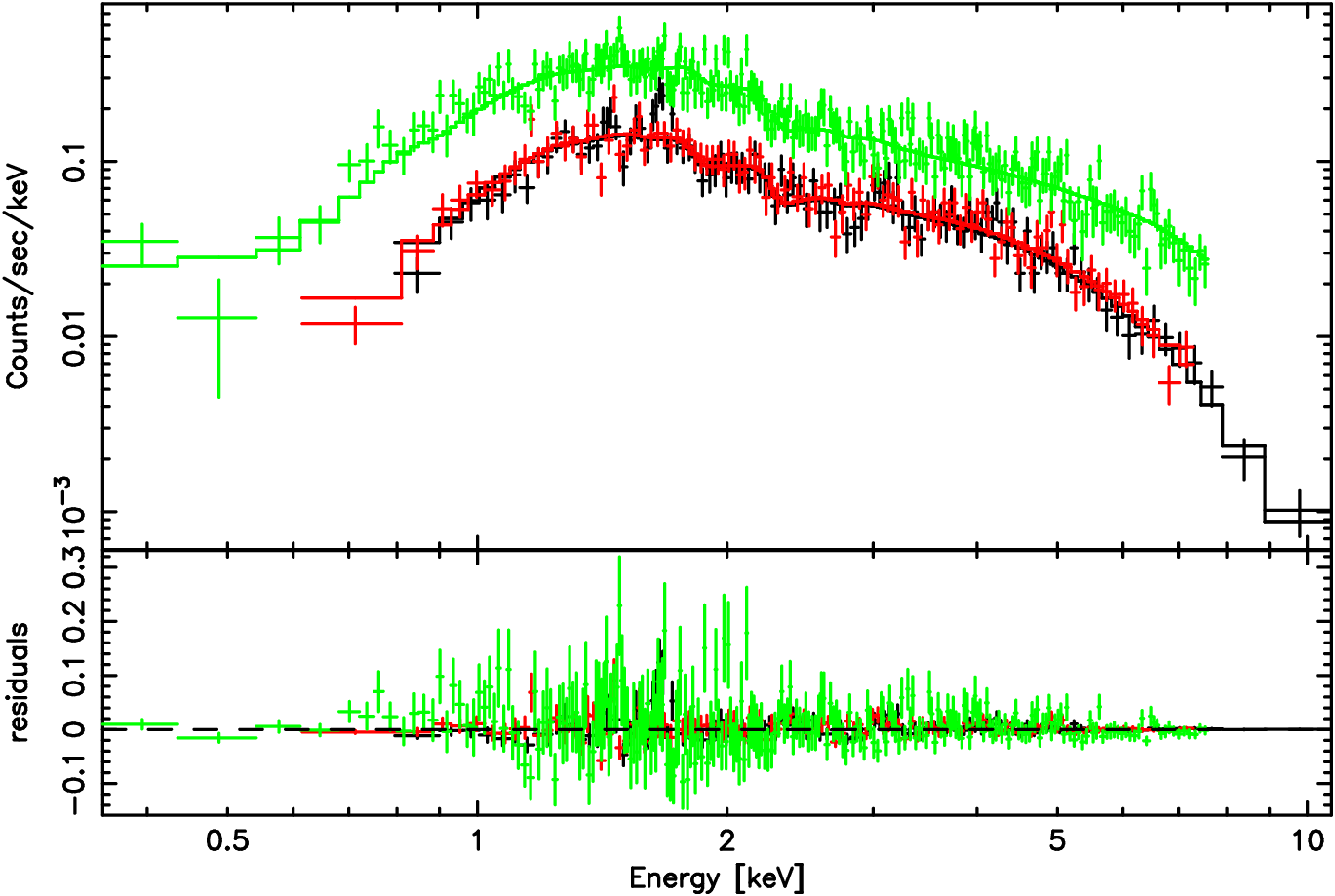}
\hfill
\caption{PN, MOS1 and MOS2 2003 March 8 (left) and March 27 (right) data of
LS\,5039\,/\,RX\,J1826.2-1450 are well fitted with an absorbed powerlaw (model I). }
\label{fig:lines}
\end{figure*}
        
Two pointings of LS\,5039 were performed by {\it XMM-Newton} in 2003:  the first one took
place on March 8 (07:33:27--10:28:38 UT),  the second on March 27 (20:53:47--23:48:59 UT).  
We analysed EPIC pn (6935 and 6981 counts), MOS1 (3979 and 3437 counts) and MOS2 
(3461 and 3521 counts) events. The pointings were performed with Medium  filters on; 
the observing
modes were Prime Small Window (time resolution $\sim 6$ ms) and Prime Partial W2, for the pn
and MOS cameras respectively. Too few counts were collected in the RGS cameras to get any
useful information.
We checked for high background phases, and chose not to set any time filtering given the very
modest disturbance (the overall MOS background flux was about 5\% of that of the source). 

We rebinned the spectrum to a minimum of 20 counts/bin and fitted the
data from the three instruments together, 
with the same models used for the {\it Chandra} data. 

A simple powerlaw provides a good fit to the combined EPIC data for both observations
(Figure~\ref{fig:lines}). 
Again, a (narrow or broad) Fe line has a null EW within the 90\% confidence level. There  is
no evidence of edges or fluorescent lines at lower energies either,
which would have given 
information on the possible cold surrounding medium (e.g. the stellar wind). On the other
hand, {\it XMM-Newton} throughput and sensitivity at the lowest
energies allow better constraints on the hydrogen column density: we
get $N_{\rm H} = 0.70\pm0.05 \times 10^{22}$ cm$^{-2}$ in powerlaw
models (I and II).   

Model III is compatible with a lower $N_{\rm H}$ value. However, this
model -- pure disk  thermal emission -- is less plausible both from
the physical and the statistical 
point of view: on one hand, it cannot account  for plasma
contributions, such as a Comptonization tail; 
on the other hand, the resulting parameters have no meaning in such low
accretion regime (from our fits the resulting innermost radius temperatures are unphysically
high). Besides, by adding a disk multi-blackbody component over the powerlaw we obtain 
negligible normalisations for the thermal emission ($L_{\rm diskbb}/L_{\rm total} < 0.09$ and
$0.20$, in the two observations respectively).  We therefore conclude that there is no
evidence of disk emission.

The addition of a simple blackbody on the top of the powerlaw
component (model IV) does not improve the fit  in a statistically significant manner, 
either,  with respect to the single powerlaw model: 
the corresponding blackbody  radius is consistent
with zero at the 90\% confidence level. By fixing the blackbody temperature at a more
realistic value -- e.g. 0.2 keV -- the best-fit upper limit on the radius of the
blackbody-emitting  surface is $\sim 1.0$ and $\sim 1.4$ km, in the two observations
respectively.

Assuming a simple powerlaw continuum (model I), the unabsorbed 2--10 keV flux is 
$F_{0.3-10 keV} \sim 10.3$ and $9.7 \times 10^{-12}$  erg cm$^{-2}$ s$^{-1}$, for 
the first and second observation respectively.  
Therefore, LS\,5039 did not vary over the 19 day time interval and  was still
a factor $\sim 4$ fainter than in the 1998 {\it Rossi-XTE} observation. Nevertheless, the slow
increase of the source X-ray luminosity  since the 2000 {\it BeppoSAX} observation seems
confirmed. 

The powerlaw index $\Gamma$  is now $\sim 1.5$,
as typical of most X-ray binaries in the hard/low state. To make a comparison with other
radio-emitting HMXBs, we note that this  value is similar to the value of $\Gamma$ in a
radio-outburst phase of LS\,I\,+61$\degr$303 (Massi 2004) and to that of Cygnus\,X-1 in 
some of the low state observations (e.g. Di Salvo \ea, 2001). \\

In order to search for pulsations, we applied the $Z_1^2$ (Rayleigh)
test (see e.g. Zavlin \ea\ 2000, and references therein) on the events collected 
by EPIC pn in a narrow circle centred on the source.  
We found no significant peaks in the $Z_1^2$ ``forest'' in the frequency range
0.001--83 Hz: in both observations, upper limits on the pulsed fraction can be set 
at $f_{\rm P} \simlt 10 \%$, whereas a rejection at the $3 \sigma$ level would have 
corresponded to $f_{\rm P} < 11.3 \%$.

\begin{figure*}[t]
\includegraphics[angle=-90,width=0.49\textwidth]{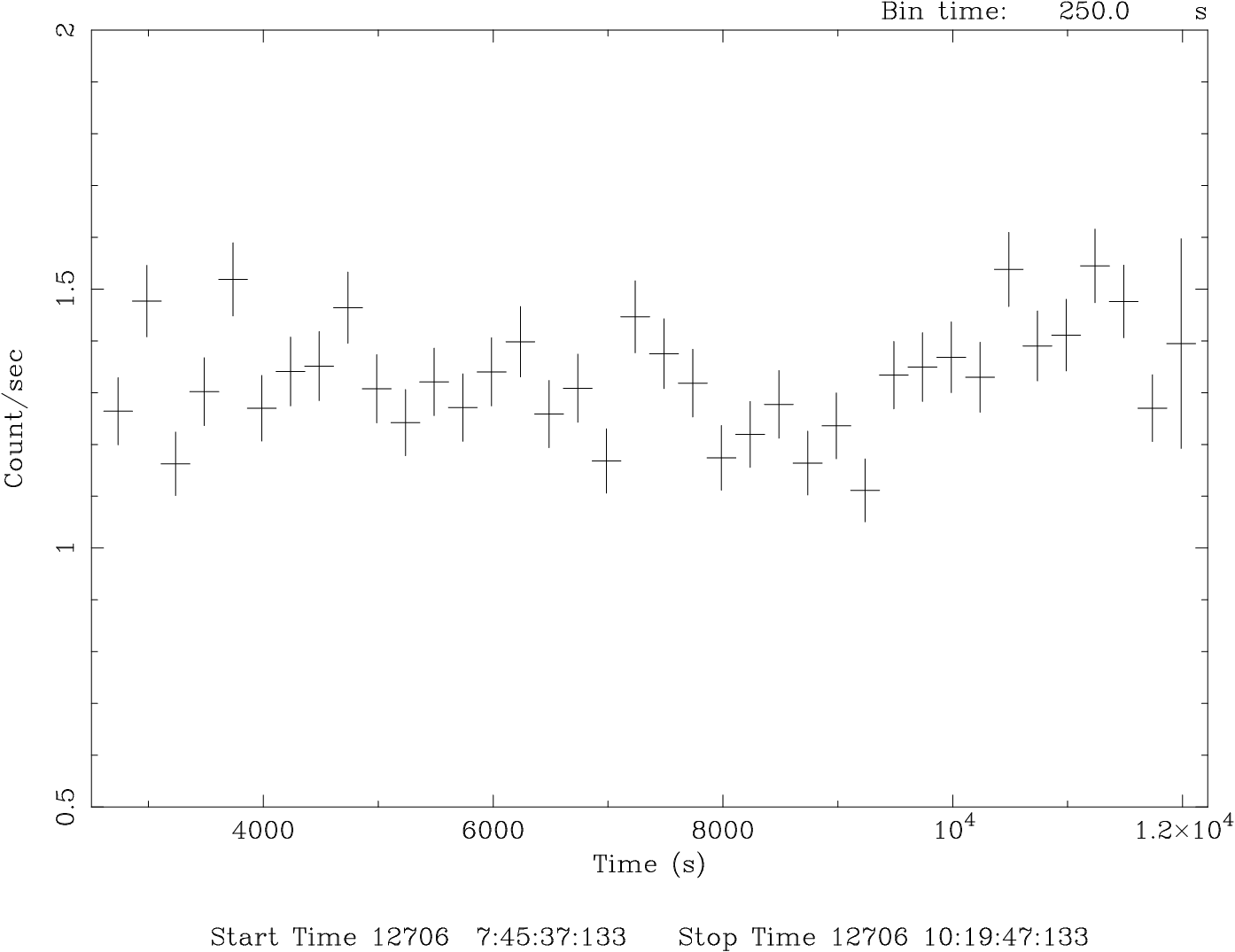}
\hfill
\includegraphics[angle=-90,width=0.49\textwidth]{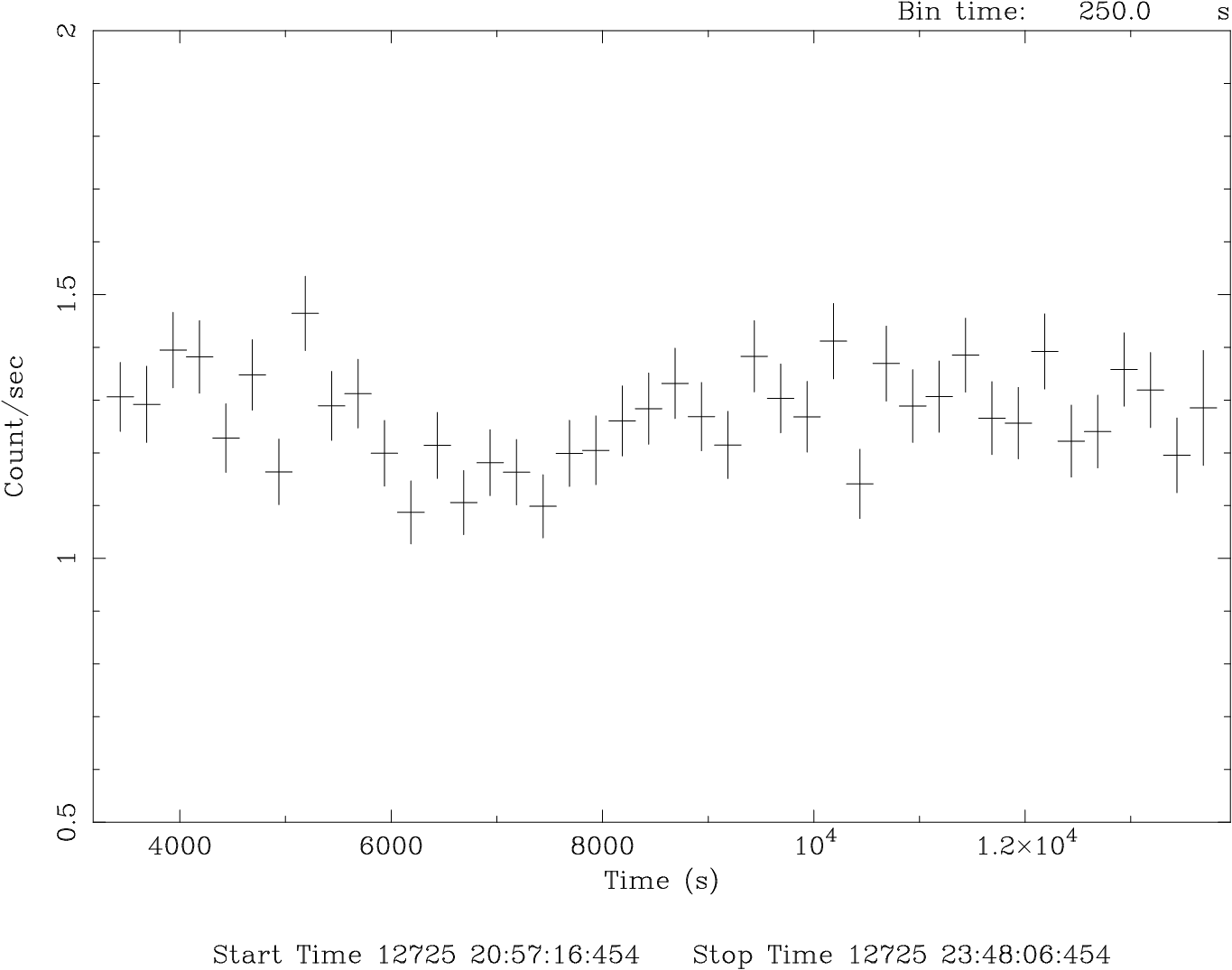}
\hfill
\caption{All-camera, 0.3--7.5 keV lightcurves of the two observations performed with 
{\it XMM-Newton}, respectively on 2003 March 8 (left) and on March 27 (right).
Notice that in correspondence of a binning time of 250 s, the maximum variation 
is at the 20\% level. }
\label{fig:varia}
\end{figure*}

As far as the short time scale variability is concerned, the lightcurves only show slight 
variation during each pointing (see Figure~\ref{fig:varia}), with no evidence of aperiodic
random fluctuations typical of wind accretors. The latter usually display flaring with
flux changes of a factor $\sim 5$ over timescales  of a few hundred seconds: we tried with
different binnings (in the range 10--1000 s) and found fluctuations of $\sim$ 20\% .

        \section{Discussion}
        \label{sec:all}

LS\,5039\,/\,RX\,J1826.2-1450 has been observed with all major X-ray telescopes since its 
identification in 1997: after its discovery with {\it ROSAT}, the source 
was found in a similar state by {\it Rossi-XTE} in 1998. At that time its luminosity was about
a factor 1000 lower than that of classical HMXBs accreting by  Roche-lobe overflow, and in the
low-end of the distribution of massive binaries fed by wind accretion. This luminosity is
comparable to those of Be/X-ray binaries in the low state, but, unlike most of them, 
LS\,5039 displays no detectable X-ray pulsations, and the primary star, identified as a 
O6.5V((f)), does not show evidence of large amount of circumstellar material in the optical.

If the association with the corresponding {\it EGRET} source is correct, the X-ray luminosity is
even smaller than the $\gamma$-ray luminosity. A similar situation occurs only in another
radio-emitting HMXB, LS\,I\,+61$\degr$303  (see Massi \ea\ 2001, Massi 2004, and references
therein). LS\,I\,+61$\degr$303 seems to contain a neutron star, although it shows no pulsations 
nor high energy cutoff up to $E \sim 25$ keV (see discussion in Greiner \& Rau, 2001). 
Its primary is a Be star with a slowly expanding equatorial  envelope
from which a transient accretion disk could form at some orbital phases. The strong emission
lines from the disk-like equatorial outflow may mask the signature of the compact object,
although a weak component moving with the orbital period has probably been  detected (Liu \ea\
2000; Zamanov \& Mart\'{\i} 2000). 

        \subsection{A wind accretor? }
        \label{sec:wind}

The orbital solution (relatively long orbital period) and 
the lack of ellipsoidal variations 
in the optical (Mart\'{\i} \ea\ 2004) exclude the presence of a Roche-lobe overflow. It is 
thus likely that accretion takes place through stellar wind. This assumption is compatible
with the low observed source X-ray luminosity  (Mc~Swain \& Gies 2002). 

LS\,5039 has been faint in the X-rays since its discovery, but has been emitting steadily in
the radio band. Classical accretion theory apparently does not provide a way
to form a large and stable accretion disk in a system with the orbital parameters of
LS\,5039 since the material in the wind carries too little angular momentum. Although small
transient discs with different rotation directions may form in such configurations, the
discs may be too small to account for the ejection process seen in radio (see e.g. Ruffert
1999). LS\,5039 could thus be a rare case in which symmetric, collimated jets form without
the help of a well formed accretion disk.

On the other hand, the non-detection of broad optical lines moving with the X-ray source
(Clark \ea\ 2001) is still no proof of the absence of a large disk, 
for instance due to the strong photospheric continuum of the star; 
and at present, the general  understanding of the processes
leading to the formation of bipolar outflows  includes the idea that jets are indeed
intimately  linked to an accretion disk around a potential well. Accretion does not  need to
be super-Eddington to produce jets. The fact that the observed jet velocity in all sorts of
physical situations (from the pre-main-sequence-stars  up to the most extreme quasars) scales
with the escape velocity suggests  that they take their origin close to the compact object
(see e.g. Livio 1997).  The persistent non-thermal radio emission of LS\,5039 is reminiscent
of the prototype  BH candidate (HMXB) Cygnus\,X-1, whose radio jet probably also originates in
the accretion  disk.

A similar, ``difficult'' case is that of SS\,433, also one of the few known radio-emitting 
HMXB (together with LS\,5039, Cyg\,X-1, Cyg\,X-3, CI\,Cam and LS\,I\,+61$\degr$303; 
but see Charles \& Coe 2003, for a more precise classification and discussion). In
this source the X-ray luminosity is $\sim 10^{35}$ erg s$^{-1}$, not far away from
that of LS\,5039. However, an accretion disk is clearly detected in the spectra of SS\,433, 
and strong red-  and
blue-shifted lines are prominent in the optical spectrum. The energy budget at the impact
point of the jet with the interstellar medium yields a jet mechanical energy of $\sim
10^{40}$ erg s$^{-1}$, clearly indicating that SS\,433 transforms most of its accretion
energy into kinetic energy and that, therefore, only a very small fraction of the accretion
power is radiated in the X-rays. \\

Long term changes in the X-ray flux appear to be 
correlated with the equivalent width of the H$\alpha$ line (Reig \ea, 2003, and McSwain \ea,
2004): this favours an explanation of the variation in terms of mean wind density changes. 
Alternatively, the explanation could lie in the fact that the X-ray luminosity 
of a NS or a BH accreting from the high velocity wind
of the primary star should vary with orbital phase. 
In fact, the entire flux range spanned by the source since its discovery 
(a factor $\sim 10$, see Table~\ref{gamma}) could be perhaps explained by 
differences in the orbital phases of the various observations (Reig \ea, 2003);
and one could note that, assuming the revised ephemeris of McSwain 
et al. (2004), the X-ray brightest {\it Rossi-XTE} observations took place at a phase 
closest to periastron passage (see Table~\ref{gamma}). However, apart from
a number of complications observed in other sources, which may question this 
interpretation,\footnote{ For instance, the fact that the orbital X-ray lightcurve of the 
wind accreting source GX\,301-2 peaks about 0.07 in phase before periastron as 
a result of wind rotation and stream flows within the system (Leahy 2002). } 
the already mentioned {\it BeppoSAX} results (Reig \ea\ 2003) when combined 
with the ephemeris of McSwain et al. (2004) are clearly 
inconsistent with the hypothesis that orbital modulation play a major role, since 
the lightcurve only shows a steady flux decrease whereas 
the flux should peak close to periastron. 

The original period of McSwain et al. (2001) used to plan the two {\it XMM-Newton} observations
predicts orbital phases of 0.95 and 0.70  for the 2003 March 8 and 27 observations
respectively, in which case a large flux variation would have been expected. However,
according to the revised ephemeris of McSwain et al. (2004), the phases of the first and
second {\it XMM-Newton} observations are now 0.79 and 0.21, i.e. at symmetric times before and
after the periastron passage, and should therefore show similar fluxes, as actually observed.
However, since other 4-d alias orbital periods may not be ruled out yet, we believe that it is
still difficult, at present, to completely distinguish between orbital and long-term 
wind variability. \\

The total absorption of LS\,5039 has been better constrained by {\it XMM-Newton}, thanks to 
the satellite's sensitivity at low energies. We obtain values around $N_{\rm H} \sim 7 
\times 10^{21}$ cm$^{-2}$. This is comparable to the values derived from optical reddening 
observations  (see Introduction), and approaches the value  estimated for the total galactic
interstellar absorption in the same direction  ($N_{\rm H} \sim 8.7 \times 10^{21}$ cm$^{-2}$:
Dickey \& Lockman 1990). Therefore, contrary to the comparable case of 4U\,1700-37
(Hammerschlag-Hensberge \ea\ 1990), no large variable local absorption is seen in LS\,5039,
at least at the time of the {\it XMM-Newton} observations.

        \subsection{The young pulsar scenario }
        \label{sec:nstar}

The nature of the compact object in LS\,5039 still remains an unsolved issue. Although the
hypothesis of a central black hole cannot be ruled out, the orbital solution rather favours a
neutron star assumption. This would not contradict the non-detection of pulsations, as shown
e.g. by the cases of 4U\,2206+54 (Torrej\'on \ea\ 2004) and 4U\,1700-37, which are HMXBs too. 
However, the latter two objects show some differences with respect to LS\,5039, including the 
fact that they are not radio sources. The only neutron star candidate, radio-emitting HMXB 
which shows no pulsations, together with LS\,5039, is precisely LS\,I\,+61$\degr$303.

Proper motion, measured in the radio and optical domain, together with  optical radial
velocities indicate that LS\,5039 is escaping its local standard  of rest with a high kick
velocity of $\sim$ 150 km\,s$^{-1}$ (Rib\'o et al.  2002). These authors conclude that its
present height above the galactic plane  is consistent with a maximum age of about
1.1\,$\times$\,10$^{6}$\,yr. If the accreting companion were a neutron star it would
therefore be a young or, at most, a  middle-aged pulsar.

Let us briefly discuss the hypothesis of a very young neutron star. It may be still rotating
too fast to accrete, and be in the propeller regime; this mechanism is still not understood
in detail and it is difficult to say whether the inferred X-ray properties would be
consistent with the observations. It is believed that such a mechanism can give rise to
fast ejection, but, considering the strong influence of the high-mass companion wind, 
it is unclear whether symmetric, well collimated jets could be generated. 

Young neutron stars have spin-down luminosities which can be comparable with the total
estimated radio plus X-ray plus $\gamma$-ray luminosity radiated by the companion of 
LS\,5039 (a few 10$^{35}$\,erg\,s$^{-1}$), and their ages are still consistent with the 
absence of existing nearby SNR (see e.g. Becker \ea\ 2002). 
The possibility thus remains that part of  the observed luminosity is
actually extracted from the relativistic wind  of a pulsar which carries away most of the
rotational energy, instead  of being due to accretion onto the compact object or on its
magnetosphere if propeller effects dominate. In the case of the binary
pulsar PSR\,1259$-$63 (spin down age $3\,\times\,10^{5}$\,yr and spin down luminosity 
of  $8\,\times\,10^{35}$\,erg\,s$^{-1}$) orbiting a Be star, Tavani and Arons (1997) 
argue that shock powered emission in a pulsar wind termination shock at a large distance 
can account for the observed X-ray luminosity of a  few 10$^{34}$\,erg\,s$^{-1}$ through 
various mechanisms (see also Murata \ea, 2003, and references therein). 

In the case of the Be system LS\,I\,+61$\degr$303,  which displays radio, X-ray and possibly
also $\gamma$-ray  luminosities similar to LS\,5039, Harrison et al. (2000) propose 
that X-ray and  $\gamma$-ray  emission arises from inverse Compton scattering of 
stellar photons on electrons accelerated in the shock between the stellar wind and the 
pulsar wind. The origin and composition of the latter (hadrons or only leptons?) should be 
investigated, as well as the conditions under which the relativistic pulsar wind would be 
actually able to create a cavity within the stellar wind.\footnote{ This may be a greater
problem in the case of Be stars, whose equatorial outflows are assumed to be denser than in 
the case of O-type stars. The effect of the orbital eccentricity would make a difference, too.} 
Such a cavity should be symmetric 
and large enough to account for the shape and size of the ejections which are observed in 
the radio band: this is an open issue, still to be adressed by proper modelling. 
The recent discovery of radio jets in LS\,I\,+61$\degr$303, with a 
geometry and size  ($\sim 0.01$ pc) quite similar to those seen in LS\,5039, has been 
considered to argue against the wind shock mechanism and for an accretion-ejection process 
(Massi et al. 2004). We note, however, that young neutron stars such as the Vela pulsar
can produce well formed jets visible in X-rays (Pavlov \ea\ 2003):  such jets are thought 
to be associated with collimated outflows of relativistic particles along the pulsar's spin 
axis. In the case of the Vela pulsar, the jet scale is of the order of one parsec, i.e.
two orders of magnitude greater than in LS\,I\,+61$\degr$303 and LS\,5039, while the
pulsar's age could be as much as two orders of magnitude less than in the latter sources. 
At a lower rotational luminosity scale, the distorted shapes of the  optical bow shock nebulae 
of PSR\,B0740-28 and PSR\,J2124-3358 rather point towards anisotropy in the pulsar wind, 
in addition to structures in the  ambient ISM (see e.g. Gaensler 2004, and references therein). 

The possible lack or weakness of any orbit-related flux variation could also hold as an 
argument against the accreting neutron star scenario.
We are thus perhaps witnessing in LS\,5039 the effects of the interaction  between the
relativistic collimated wind of a young pulsar with the O star  wind. It is beyond the scope
of this paper to discuss the shock structures and emission mechanism which could account for
the  radio to $\gamma$-ray energy distribution, but we wish, however, to question in this
particular case the paradigm which systematically associates bipolar outflows with accretion
discs. 

\begin{table*}[htbp]
\centering
\begin{tabular}{|c|ccc|}
\hline
&                          &                              &                              \cr
& {\it Chandra}, 10/09/02  & {\it XMM-Newton}, 08/03/03 & {\it XMM-Newton}, 27/03/03 \cr
&                          &                              &                              \cr
\hline
&                          &                              &                     \cr
actual duration [s] 
& $10608$          & $\sim 10650$ (MOS), $10510$ (pn) & $\sim 10665$ (MOS), $10510$ (pn) \cr
count rate [s$^{-1}$] 
& $0.094 \pm 0.003$        & $0.919 \pm 0.012$ (pn)       & $0.878 \pm 0.012$ (pn) \cr
&                          &                              &                     \cr
\hline
model I & $61.7/44 = 1.40$ & $465.7/478 = 0.97$           & $510.0/460 = 1.11$   \cr
\hline
&                          &                              &         \cr
$N_{\rm H}$ [$10^{22}$ cm$^{-2}$]
& $0.56^{+0.35}_{-0.19}$   & $0.72^{+0.03}_{-0.05}$       & $0.69^{+0.05}_{-0.03}$ \cr
$\Gamma$            
& $1.15^{+0.23}_{-0.21}$   & $1.56^{+0.02}_{-0.05}$       & $1.49^{+0.05}_{-0.04}$ \cr
&                          &                              & \cr
\hline
model IIa (narrow line) & $61.7/43 = 1.43$ & $465.4/477 = 0.98$ & $510.8/459 =1.11$  \cr
\hline
&                          &                              &         \cr
$N_{\rm H}$ [$10^{22}$ cm$^{-2}$]
& $0.58^{+0.35}_{-0.29}$   & $0.69^{+0.05}_{-0.02}$       & $0.70^{+0.05}_{-0.03}$  \cr
$\Gamma$            
& $1.17^{+0.23}_{-0.22}$   & $1.53^{+0.04}_{-0.04}$       & $1.50^{+0.03}_{-0.05}$  \cr
&                          &                              & \cr
$E_0$(Fe K$\alpha$) [keV] 
& $6.4$ (fixed)            & $6.4$ (fixed)                & $6.4$ (fixed) \cr
EW(Fe K$\alpha$) [eV]
& $5^{+288}_{-5}$          & $0 (<<1)$                    & $0 (<87)$     \cr 
$\sigma$(Fe K$\alpha$) [eV] 
& $20$ (fixed)             & $20$ (fixed)                 & $20$ (fixed)       \cr
&                          &                              &         \cr
\hline
model IIb (broad line) & $61.5/43 = 1.43$ & $465.5/477 = 0.98$ & $510.8/459 =1.11$  \cr
\hline
&                          &                              &         \cr
$N_{\rm H}$ [$10^{22}$ cm$^{-2}$]
& $0.60^{+0.40}_{-0.32}$   & $0.69^{+0.05}_{-0.02}$       & $0.70^{+0.05}_{-0.04}$  \cr
$\Gamma$            
& $1.20^{+0.27}_{-0.26}$   & $1.53^{+0.04}_{-0.04}$       & $1.50^{+0.04}_{-0.04}$  \cr
&                          &                              & \cr
$E_0$(Fe K$\alpha$) [keV] 
& $6.4$ (fixed)            & $6.4$ (fixed)                & $6.4$ (fixed) \cr
EW(Fe K$\alpha$) [eV]
& $165^{+564}_{-165}$      & $0 (<137)$                    & $0 (<187)$ \cr 
$\sigma$(Fe K$\alpha$) [eV] 
& $390$ (fixed)            & $390$ (fixed)                & $390$ (fixed)       \cr
&                          &                              & \cr
\hline
model IIIa (narrow line) & $56.9/42 = 1.35$ & $566.0/477 = 1.19$ & $539.4/459 = 1.18$  \cr
\hline
&                          &                              &  \cr
$N_{\rm H}$ [$10^{22}$ cm$^{-2}$]
& $0.28^{+0.16}_{-0.14}$   & $0.44^{+0.03}_{-0.02}$       & $0.45^{+0.04}_{-0.02}$ \cr
$T$ (blackbody) [keV] 
& $3.85^{+0.77}_{-0.67}$   & $2.34^{+0.08}_{-0.09}$       & $2.29^{+0.09}_{-0.09}$ \cr
&               &                    & \cr
$E_0$(Fe K$\alpha$) [keV] 
& $7.04^{+0.20}_{-1.67}$   & $6.4$ (fixed)                & $6.4$ (fixed) \cr
EW(Fe K$\alpha$) 
& $331^{+E13}_{-331}$      & $0 (<82)$                    & $0 (<97)$   \cr 
$\sigma$(Fe K$\alpha$) [eV] 
& $20$ (fixed)             & $20$ (fixed)                 & $20$ (fixed)       \cr
&                          &                              &         \cr
\hline
model IIIb (broad line) & $58.4/42 = 1.39$ & $566.0/477 = 1.19$ & $537.0/459 = 1.17$  \cr
\hline
&                          &                              &  \cr
$N_{\rm H}$ [$10^{22}$ cm$^{-2}$]
& $0.32^{+0.16}_{-0.14}$   & $0.44^{+0.03}_{-0.02}$       & $0.48^{+0.02}_{-0.08}$ \cr
$T$ (blackbody) [keV] 
& $3.39^{+0.89}_{-0.55}$   & $2.34^{+0.08}_{-0.09}$       & $2.18^{+0.09}_{-0.10}$ \cr
&                          &                              & \cr
$E_0$(Fe K$\alpha$) [keV] 
& $7.09^{+1.01}_{-0.53}$   & $6.4$ (fixed)                & $6.4$ (fixed) \cr
EW(Fe K$\alpha$) 
& $699^{+E13}_{-699}$      & $0 (<149)$                    & $0 (<472)$   \cr 
$\sigma$(Fe K$\alpha$) [eV] 
& $390$ (fixed)            & $390$ (fixed)                & $390$ (fixed)    \cr
&                          &                              &                  \cr
\hline
model IV & --              & $460.2/476 = 0.97$           & $503.5/458 = 1.10$  \cr
\hline
&                          &                              &         \cr
$N_{\rm H}$ [$10^{22}$ cm$^{-2}$]
& --                       & $0.67^{+0.04}_{-0.08}$       & $0.66^{+0.10}_{-0.05}$ \cr
$\Gamma$            
& --                       & $1.53^{+0.04}_{-0.04}$       & $1.52^{+0.15}_{-0.06}$ \cr
$T$ (blackbody) [keV] 
& --                       & $0.95^{+0.13}_{-0.19}$       & $1.14^{+0.09}_{-0.16}$ \cr
$r$ (blackbody) [km]
& --                       & $0 (<0.07)$                  & $0 (<0.05)$ \cr
&                          &                              &    \cr
\hline
\end{tabular}
\caption{Best-fit parameters for the spectral models listed at the
end of Section 2. $\chi^2/dof$ values are given in correspondence of
each model's name and for each observation. }
\label{tab:fits}
\end{table*}

        \subsection{What about the iron line? }
        \label{sec:line}

The detection by {\it Rossi-XTE} of a strong, broad fluorescent iron
line (Rib\'o \ea\ 1999)  cannot 
be confirmed with the better resolution X-ray data provided by {\it
Chandra} and {\it 
XMM-Newton}, and this is apparently not just due to poor statistics in such faint states. The
broad line component ($\sigma_{\rm line}=390$ eV) claimed by  Rib\'o \ea (1999)  has null
equivalent width even within the 99\% confidence level in {\it XMM-Newton} data, with an
upper limit of 0.14 keV compared to the 0.75 keV observed by Rib\'o \ea\ (1999). F-tests lead
to exclude all broad line as well as narrow line models (IIa,b and IIIa,b) with high
confidence. Let us shortly comment on this.

The detection and the study of fluorescent iron emission in accreting galactic BH 
candidates can be a powerful tool to test assumptions on the extreme gravitational field as 
well as on the accretion/ejection flows which are present in the innermost parts of such  
systems, at just a few gravitational radii ($r_{\rm g}=GM/c^2$) from the BH horizon or compact object
surface.\footnote{See e.g. Reynolds \& Nowak (2002) for a review  on fluorescent iron lines as
a probe of astrophysical BH systems, and Martocchia,  Karas \& Matt (2000) on details about
the Fe line diagnostics in the ``cold'' disk  assumption. } This was first demonstrated by the
case of some Active Galactic Nuclei  (e.g. Tanaka \ea\ 1995), but soon observational campaigns
of galactic BH candidates, performed with all recent X-ray satellites, showed that
broad, intense  Fe lines are not rare in stellar-mass sources. However, such phenomenon
generally occurs  in ``intermediate / very high'' spectral states, while after 1999
LS\,5039\,/\,RX\,J1826.2-1450 has been observed  always in a ``low/hard'' state. 

Broad lines, best-fitted with a relativistic disk model, have been observed 
among others in Cygnus\,X-1, 
XTE\,J1650-500, V4641\,Sgr, XTE\,J2012+381. In the cases of XTE\,J1650-500 and GX\,339-4, {\it
XMM-Newton} detected very broad, skewed  lines, best-fitted with a canonical Kerr innermost
stable orbit, and a very  steep emissivity  profile (Miller \ea\ 2002 and 2004). GRS\,1915+105,
too, showed a broad,  distorted iron line in a few {\it BeppoSAX} observations (Martocchia
\ea, 2002 and 2004). In other sources, the lines and their profiles are rather interpreted as
products of  ejected material: besides XTE\,J1550-564 and 4U\,1630-47, this is the case of 
SS\,433, where two (red- and blue-shifted) iron line components are visible, which follow the
same  velocity variations of several other emission lines, observed with ASCA SIS and Chandra
HETGS  (see Kotani \ea\ 1994 and Marshall \ea\ 2002, respectively). A broad, relatively 
intense fluorescent iron line has been observed in CI\,Cam as well, with various instruments 
including {\it XMM} (e.g. Ueda \ea\ 1998, Boirin \ea\ 2002). \\

The line observed with the PCA onboard {\it Rossi-XTE}, if confirmed by an X-ray observation
of the source in a brighter state, could have a disk or a jet origin.  In the first case, the
line may carry the general-relativistic imprints  of the strong gravitational potential well
in the vicinity of compact objects; and the blue-shift -- 6.6 instead of the 6.4 keV iron
K$\alpha$ rest energy -- would imply a substantial disk ionisation. In the disk assumption,
the fact that the line is not  seen in the faintest states could be a consequence of the disk
disappearance in the  ``jet'' (low/hard) state. On the other hand, in galactic stellar-mass
sources the  ionisation of the accreting flow is thought to be much more relevant than in
Active  Galactic Nuclei because of the much larger disk temperature in stellar size objects; 
this sometimes results in fluorescent emission being less effective. Moreover, the 
non-negligible X-ray variability of such sources, which reflects rapid changes in the  physics
and geometry of the accretion/ejection phases, also causes the iron emission to be non-steady,
and even non-detectable in many situations. 

If emitted from a jet, instead, the line would be shifted in  energy mainly as a
consequence of the Doppler effect, due to the bulk velocity of the jet material, and the
broadening may be interpreted as a signature of the jet opening angle. The jet matter, on its
account, must be ``cold'' to emit the line efficiently. Since in the  case of LS\,5039 the
radio-observed ejections are symmetric (Paredes \ea\  2000, 2002), the issue would 
be to self-consistently explain why the line is observed in only one of the two jets. The
fact that this is the jet approaching the observer leads us to invoke Doppler boosting. 

In absence of pulsations, lacking a precise determination of the compact object's mass, and 
since we have no indications on the presence or absence of a high-energy cutoff, the iron 
line diagnostics would be crucial to investigate the nature of the  compact
object.  A dedicated {\it XMM-Newton} observation to catch the source in a bright, Fe 
line-emitting state was proposed also for AO3. The idea is  to trigger the observation by
optical monitoring, using the fact that an anticorrelation between the measured
(negative) EW(H$\alpha$) and the observed X-ray flux -- as claimed by Reig \ea\ (2003) -- is
likely due to the  connection between the primary mass-loss rate (which governs the optical
emission)  and the mass accretion rate onto the compact object (which governs the X-rays). 
However, LS\,5039\,/\,RX\,J1826.2-1450 has failed to enter such a ``bright" state, requested
as a  triggering condition, up to now.

        \section{Conclusions}
        \label{sec:concl}

Recent {\it XMM-Newton} and {\it Chandra} observations of LS\,5039\,/\,RX\,J1826.2-1450 
caught the source in a faint state.  The X-ray spectrum can be fitted by an absorbed 
powerlaw, without the need for any disk component. Simple blackbody
emission, which could be generated at a neutron star's surface, is ruled out, too. The
powerlaw shape is hard ($\Gamma = 1.1\div1.5$), similar to that of most X-ray 
binaries in the low/hard state. 

In contrast to previous {\it RXTE} observations, we do not find any evidence of iron line
emission. Even if the iron line in LS\,5039\,/\,RX\,J1826.2-1450 is a transient
feature, data from satellites with better sensitivity and resolution, observing the source in
a brighter state, would be necessary to confirm or reject the {\it Rossi-XTE} claim.

{\it XMM-Newton} data allow a better constraint on the hydrogen column
density ($N_{\rm H} \sim 7 \times 10^{21}$ cm$^{-2}$), excluding the
presence of large amounts of local absorption at the time of these observations.

The absence of pulsations down to low pulse fractions is confirmed. We also 
fail to detect the flaring activity often seen in wind accreting systems: the X-ray 
flux of the source varies by just $\sim$ 20\% on timescales of a few hundred seconds within 
each {\it XMM-Newton} observation.

The overall flux history of the source is summarized in Table~\ref{gamma}.
While a general correlation is possible, and can be seen (although not so
strict) between spectral ``hardness'' and X-ray flux, the latter shows no clear 
dependence on the orbital phase, assuming the ephemeris by McSwain \ea\ (2004). 
Although part of the long term flux variations still
could be due to orbital effects in a wind accretion scenario, many other arguments 
rather favour an interpretation not based on accretion. 
They include: the likely physical impossibility of forming a large 
disk with the given orbital parameters; the non-detection of disk spectral features; 
the apparent lack of intrinsic absorption, and the non-detection of edges or fluorescent 
lines in the soft X-rays, which would have given information on the stellar wind.

We therefore discussed the hypothesis that LS\,5039/RX\,J1826.2-1450 may be a young pulsar, 
and that X-ray emission may arise from the interaction between its relativistic
collimated wind and the wind of the primary star. This would break the paradigm which 
systematically associates bipolar outflows with the presence of an accretion disc. 
X-ray observations spanning a complete orbital period, possibly accompanied by
optical monitoring of the H$\alpha$ feature, are strongly needed to finally discriminate 
between the various models for the accretion/ejection mechanism as well as for the 
nature of the compact object in LS\,5039/RX\,J1826.2-1450.

\begin{acknowledgements}
The authors are grateful to the anonymous referee, to Marc Rib\'o and to Slava Zavlin 
for their useful comments and suggestions. AM also wishes to thank Giorgio Matt for the 
help, and CNES for financial support. 
IN is a researcher of the programme {\em Ram\'on y Cajal}, funded by the Spanish Ministerio 
de Ciencia y Tecnolog\'{\i}a and the University of Alicante.
This research is partially supported by the Spanish Ministerio de
Ciencia y Tecnolog\'{\i}a under grant ESP2002-04124-C03-03.
\end{acknowledgements}

\end{document}